\documentclass[copyright,creativecommons]{eptcs}
\providecommand{\event}{GandALF 2013} % Name of the event you are submitting to
\usepackage{breakurl}             % Not needed if you use pdflatex only.
\usepackage[strings]{underscore}

\newtheorem{definition}{Definition}
\newtheorem{proposition}{Proposition}
\newtheorem{theorem}{Theorem}
\newtheorem{lemma}{Lemma}

\usepackage{alltt}
\usepackage{graphicx}
\usepackage{color}
\usepackage{nicefrac}
\usepackage{subfigure}
\usepackage{amsmath}
\usepackage{amssymb}
\usepackage{paralist}
\usepackage{float}
\usepackage{framed}
\usepackage{smallsec}
\usepackage{stmaryrd} %\llbracket \rrbracket

\usepackage{tikz}
\usepackage{tkz-graph}

\newcommand{\qed}{\quad\hfill\mbox{$\Box$}\par\medskip}
\newcommand{\ourproof}[1]{\noindent {\bf Proof {#1}:}\ }

\newcommand{\hide}[1]{}
\newcommand{\Inf}[1]{\mathsf{Inf}({#1})}

\newcommand{\Nat}{\mathbb{N}}
\newcommand{\ccc}{{\tt {rabin}}}
\newcommand{\ccca}{{\tt {Rabin}}_a}
\newcommand{\pop}{{\tt {pop}}}
\newcommand{\cycle}{{\tt{cycle}}}
\newcommand{\getanchor}{{\tt{getAnchor}}}
\newcommand{\cccai}{\ccc^\alpha}

\newcommand{\msr}[1]{{\mu({#1})}}
\newcommand{\rabinindex}[1]{{\mathsf{RI}}({#1})}
\newcommand{\rabinindexai}[1]{{\mathsf{RI}^\alpha}({#1})}

\newcommand{\equivai}{\equiv^\alpha}

\graphicspath{{./figures/}}
 
\title{The Rabin index of parity games}

\author{Michael Huth and Jim Huan-Pu Kuo
\institute{Department of Computing, Imperial College London} \\
\institute{London, SW7 2AZ, United Kingdom} 
\email{\{m.huth, jimhkuo\}@imperial.ac.uk}
\and
Nir Piterman
\institute{Department of Computer Science, University of Leicester} \\
\institute{Leicester, LE1 7RH, United Kingdom}
\email{nir.piterman@leicester.ac.uk}
}

\def\titlerunning{The Rabin index of parity games}
\def\authorrunning{M.\ Huth, J.\ H.\ Kuo \& N.\ Piterman}

\begin{document}

\maketitle

\begin{abstract}
We study the descriptive complexity of parity games by 
taking into account the coloring of their game graphs whilst
ignoring their ownership structure. Colored game graphs are identified
if they determine the same winning regions and strategies, for
\emph{all} ownership structures of nodes. 
The Rabin index of a parity game is the minimum of the maximal color
taken over all equivalent coloring functions. 
We show that deciding whether the Rabin index is at least $k$ is in P
for $k=1$ but NP-hard for all \emph{fixed} $k\geq 2$. 
We present an
EXPTIME algorithm that computes the Rabin index by simplifying its
input coloring function. 
When replacing simple cycle with cycle detection in that algorithm,
its output over-approximates the Rabin index in polynomial
time. Experimental results show that this approximation yields good
values in practice. 
\end{abstract}

\section{Introduction}\label{sec:introduction}
Parity games (see e.g.\ \cite{Zie98}) are infinite, $2$-person,
$0$-sum, graph-based games that are hard to solve. 
Their nodes are colored with natural numbers, controlled by different
players, and the winning condition of plays depends on the minimal
color occurring in cycles. 
The condition for winning a node, therefore, is an alternation of
existential and universal quantification. 
In practice, this means that the maximal color of its coloring
function is the only exponential source for the worst-case complexity
of most parity game solvers, e.g.\ for those in
\cite{Zie98,Jur00,VJ00}.

One approach taken in analyzing the complexity of parity games, and
in so hopefully improving the complexity of their solution, is
through the study of the descriptive complexity of their underlying
game graph.
This method therefore ignores the ownership structure on parity games.

An example of this approach is
the notion of DAG-width in \cite{BDHK06}. Every directed graph
has a DAG-width, a natural number that specifies how well that graph
can be decomposed into a directed acyclic graph (DAG). The decision
problem for DAG-width, whether the DAG-width of a directed graph is at
most $k$, is NP-complete in $k$ \cite{BDHK06}. But parity games whose
DAG-width is below a given threshold have polynomial-time solutions
\cite{BDHK06}. 
The latter is a non-trivial result since DAG-width also ignores the
colors of a parity game.

In this paper we want to develop a similar measure of the descriptive
complexity of parity games, their \emph{Rabin index}, a natural number
that ignores the ownership of nodes,
but does take into account the colors of a parity game.
Intuitively, the Rabin index is the number of colors that are
\emph{required} to capture the complexity of the game structure.
By measuring and reducing the number of colors we hope to improve
the complexity of analyzing parity games. %
\footnote{
We note that if we also were to account for ownership, we could solve
the parity game and assign color $0$ to nodes won by player~0 and
color $1$ to nodes won by player~1.
Thus, this would reduce the index of \emph{all} games to at
most $2$. However, this would prevent a more fine-grained analysis of
the structural complexity of the game and defeats the purpose
of simplifying parity games \emph{before} solving them.
}
The reductions we propose are related to priority compression and propagation in
\cite{Friedman09} but, in contrast, exploit the \emph{cyclic} structure of game
graphs.

The name for the measure developed here is inspired by related work on
the Wagner hierarchy for automata on 
infinite words \cite{Wagner79}: Carton and Maceiras use similar ideas
to compute and 
minimize the Rabin index of deterministic parity automata on infinite 
words \cite{CartonM99}. To the best of our knowledge, our work is the
first to study this notion in the realm of infinite, 2-person games.

The idea behind our Rabin index is that one may change the coloring function of a parity game
to another one if that change neither
affects the winning regions nor the choices of winning strategies. 
This yields an equivalence relation between coloring functions.
For the coloring function of a parity game,
we then seek an equivalent coloring function with the
smallest possible maximal color, and call that minimal maximum the
Rabin index of the respective parity game.

The results we report here about this Rabin index are similar in spirit to those developed for
DAG-width in \cite{BDHK06} but there are important differences:

\begin{compactitem}
\item We propose a
measure of descriptive complexity that is closer to the structure of the
parity game as it only forgets ownership of nodes and not their colors. 

\item We prove that for every \emph{fixed} $k\geq 2$, deciding whether the
  Rabin index of a parity game is at least $k$ is NP-hard.

\item We can
characterize the above equivalence relation in terms of the parities
of minimal colors on \emph{simple} cycles in the game graph.

\item  We use that characterization to design an algorithm that
  computes the Rabin index and a witnessing coloring function in exponential time.

\item We show how the same algorithm efficiently computes sound
  approximations of the Rabin index when simple cycles are abstracted
  by cycles.

\item We derive from that approximation an abstract Rabin index of parity games such that games with bounded abstract Rabin index are efficiently solvable.

\item We conduct detailed experimental studies that corroborate the
  utility of that approximation, also as a preprocessor for solvers.
\end{compactitem}

\paragraph{Outline of paper.} Section~\ref{section:background}
contains background for our technical develeopments. In Section~\ref{section:equivalance},
we define the equivalence between coloring functions, characterize it
in terms of simple cycles, and use that characterization to define the Rabin index of
parity games. In
Section~\ref{section:algorithm} we develop an algorithm that runs in
exponential time and computes a coloring function which witnesses the
Rabin index of the input coloring function. The complexity of the natural decision problems for the
Rabin index is studied in Section~\ref{section:complexity}. 
An abstract version of our algorithm is shown to soundly approximate that coloring function and Rabin index in Section~\ref{section:static}. Section~\ref{section:experiments} contains our experimental results for this abstraction. 
And we conclude the paper in Section~\ref{section:conclusions}. An
appendix contains selected proofs.

\section{Background}
\label{section:background}
We write $\Nat$ for the set $\{0,1,\dots\}$ of natural numbers. A parity game $G$ is a tuple $(V,V_0,V_1,E,c)$ where $V$ is a non-empty 
set of nodes partitioned into possibly empty node sets $V_0$ and $V_1$, with an edge relation $E\subseteq V\times V$ (where for all $v$ in $V$ there is a $w$ in $V$ with $(v,w)$ in $E$), and a coloring function $c\colon V\to \Nat$.

Throughout, we write $s$ for one of $0$ or $1$. In a parity game,
player $s$ owns 
the nodes in $V_s$. A play from some node $v_0$ results in an
infinite play $P = v_0v_1\dots$ in $(V,E)$ where the player who owns $v_i$
chooses the successor $v_{i+1}$ such that $(v_i,v_{i+1})$ is in $E$.  Let $\Inf P$ be the set of colors
that occur in $P$ infinitely often: $\Inf P = \{ k\in\Nat \mid
\forall j\in \Nat \colon \exists i\in \Nat \colon i>j \mbox{ and } k=c(v_i)
\}
%\exists I\subseteq \Nat\hbox{ infinite}\colon \forall i\in I\colon k =
%c(v_i)\}
$. Player $0$ wins play $P$ iff $\min \Inf P$ is even; otherwise
player $1$ wins play $P$.

A strategy for player $s$ is a total function $\tau\colon V_s\to V$
such that $(v,\tau(v))$ is in $E$ for all $v\in V_s$. A play $P$ is
consistent with $\tau$ if each node $v_i$ in $P$ owned by player $s$
satisfies $v_{i+1} = \tau(v_i)$. 
It is well known that each parity game is determined:
node set $V$ is the disjoint union of two sets $W_0$ and $W_1$, the
winning regions of players $0$ and $1$ (respectively), where one of $W_0$ and $W_1$ may be empty. 
Moreover, 
strategies  $\sigma\colon V_0\to V$
and $\pi\colon V_1\to V$ can be computed such that
\begin{compactitem}
\item all plays
beginning in $W_0$ and consistent with $\sigma$ are won by player $0$;
and 

\item all plays beginning in $W_1$ and consistent with $\pi$ are
won by player $1$.
\end{compactitem}

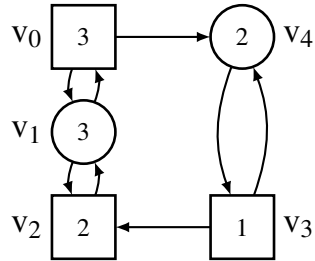
\begin{figure}[t]
\begin{center}
\begin{tikzpicture}[scale=1.2, transform shape]
\Vertex[x=0.0pt,y=0.0pt,L=v$_{0}$,LabelOut=true,Ldist=0pt,Lpos=180,style={color=white,text=black}]{v0ext}
\Vertex[x=0.0pt,y=0.0pt,L=$3$,style={font=\scriptsize, shape=rectangle,minimum height=20pt, minimum width=20pt}]{v0}

\Vertex[x=0.0pt,y=-30.0pt,L=v$_{1}$,LabelOut=true,Ldist=0pt,Lpos=180,style={color=white,text=black}]{v1ext}
\Vertex[x=0.0pt,y=-30.0pt,L=$3$,style={font=\scriptsize, minimum size=20pt}]{v1}

\Vertex[x=0.0pt,y=-60.0pt,L=v$_{2}$,LabelOut=true,Ldist=0pt,Lpos=180,style={color=white,text=black}]{v2ext}
\Vertex[x=0.0pt,y=-60.0pt,L=$2$,style={font=\scriptsize, shape=rectangle,minimum height=20pt, minimum width=20pt}]{v2}

\Vertex[x=50.0pt,y=-60.0pt,L=v$_{3}$,LabelOut=true,Ldist=0pt,Lpos=0,style={color=white,text=black}]{v3ext}
\Vertex[x=50.0pt,y=-60.0pt,L=$1$,style={font=\scriptsize, shape=rectangle,minimum height=20pt, minimum width=20pt}]{v3}

\Vertex[x=50.0pt,y=0.0pt,L=v$_{4}$,LabelOut=true,Ldist=0pt,Lpos=0,style={color=white,text=black}]{v4ext}
\Vertex[x=50.0pt,y=0.0pt,L=$2$,style={font=\scriptsize, minimum size=20pt}]{v4}

\Edge[style={->,>=latex}](v0)(v4)
\Edge[style={->,>=latex}](v3)(v2)
\tikzstyle{EdgeStyle}=[bend right =20]
\Edge[style={->,>=latex}](v0)(v1)
\Edge[style={->,>=latex}](v1)(v0)
\Edge[style={->,>=latex}](v1)(v2)
\Edge[style={->,>=latex}](v2)(v1)
\Edge[style={->,>=latex}](v3)(v4)
\Edge[style={->,>=latex}](v4)(v3)
\end{tikzpicture}
\end{center}
\vspace*{-5mm}
\caption{A parity game with winning regions $W_0 = \{v_1,v_2\}$ and
  $W_1 = \{v_0,v_3,v_4\}$; winning strategies for players $0$ and $1$ map $v_1$ to $v_2$, respectively $v_0$ and $v_3$ to $v_4$
\label{fig:pg}}
\vspace*{-5mm}
\end{figure}

Solving a parity game means computing such data $(W_0,W_1,\sigma,\pi)$.
We show a parity game and one of its
possible solutions in Figure~\ref{fig:pg}.

\section{Rabin Index}
\label{section:equivalance}
We now formalize the definition of equivalence for coloring functions,
and then use that notion in order to formally define the Rabin index
of a parity game.

We want to reduce the complexity of a coloring function
$c$ in a parity game $(V,V_0,V_1,E,c)$ by transforming $c$ to some
coloring function $c'$. 
Since we do not want the transformation to be based on a solution of the 
game we design the transformation to ignore ownership of nodes.
That is, it 
needs to be sound  for \emph{every possible} ownership
structure $V = V_0 \cup V_1$. Therefore, for \emph{all} such partitions $V = V_0\cup V_1$,
the two parity games  $(V,V_0,V_1,E,c)$ and $(V,V_0,V_1,E,c')$ 
that differ
only in colors need to be
equivalent in that they have the same winning regions and the same
sets of winning strategies. 
We formalize this notion. 

\begin{definition}
Let $(V,E)$ be a directed graph and $c,c'\colon V\to \Nat$ two
coloring functions. 
We say that $c$ and $c'$ are equivalent, written $c\equiv c'$, iff 
for all partitions $V_0\cup V_1$ of $V$ the resulting parity games
$(V,V_0,V_1,E,c)$ and $(V,V_0,V_1,E,c')$ have the same winning regions
and the same sets of winning strategies for both players.
\end{definition}

Intuitively, changing coloring function $c$ to $c'$ with $c\equiv c'$ is sound:
regardless of what the actual partition of $V$ is, we know that this
change will neither affect the winning regions nor the choice of their
supporting winning strategies. But the definition of $\equiv$ is not
immediately amenable to algorithmic simplification of $c$ to some
$c'$. This definition quantifies over exponentially many partitions,
and for each such partition it insists that certain sets of strategies
be equal.  

We need a more compact characterization of $\equiv$ as the basis for
designing a static analysis. To that end, we require some concepts
from graph theory first. 

\begin{definition}
\begin{compactenum}
\item A path $P$ in a directed graph $(V,E)$ is a sequence
  $v_0,v_1,\ldots,v_n$ of nodes in $V$
such that $(v_i,v_{i+1})$ is in $E$ for every $i$ in
  $\{0,1,\ldots, n-1\}$.
\item A cycle $C$ in a directed graph $(V,E)$ is a path
  $v_0,\dots, v_n$ with $(v_n,v_0)$ in $E$.
\item A simple cycle $C$ in a directed graph $(V,E)$ is a cycle
  $v_0,v_1,\dots, v_n$ such that for every $i \not= j$ in $\{0,1,\dots n\}$ we
  have $v_i\neq v_j$.
\item For $(V,E,c)$, the $c$-color of a cycle $v_0,\ldots, v_n$ in
  $(V,E)$ is $\min_{0\leq i\leq n} c(v_i)$.
\end{compactenum}
\end{definition}

Simple cycles are paths that loop so that no node has more than one
outgoing edge on that path. 
A cycle is defined similarly, except that it is allowed
that $v_i$ equals $v_j$ for some $i\not=j$, so a node on that path may
have more
than one outgoing edge. The color of a cycle
is the minimal color that occurs on it.

For example, for the parity game in Figure~\ref{fig:pg}, a simple cycle is
$v_0,v_4,v_3,v_2,v_1$ and its color is $1$, a cycle that is not simple
is $v_0,v_1,v_2,v_1$ and its color is $2$.

We can now characterize $\equiv$ in terms of colors of simple
cycles. Crucially, we make use of the fact that parity games have
pure, positional strategies \cite{EJ91}.

\begin{proposition}
\label{prop:equiv}
Let $(V,E)$ be a directed graph and $c,c'\colon V\to \Nat$ two
coloring functions. Then $c\equiv c'$ iff for all simple cycles $C$ in
$(V,E)$, the $c$-color of $C$ has the same parity as the $c'$-color of
$C$. 
\end{proposition}

\ourproof {Sketch}
We write $c\sim c'$ iff for all simple cycles $C$ in $(V,E)$, the $c$-color of $C$ has the same parity as the $c'$-color of $C$. We have to show $\sim$ equals $\equiv$.

To prove that $\sim$ is contained in $\equiv$, let $c\sim c'$
be given. For each subset $V_0$ of $V$ we have parity games
$G_c = (V,V_0,V\setminus V_0,c)$ and $G_{c'} = (V,V_0,V\setminus
V_0,c')$. We write $W_s$ (resp.\ $W_s'$) for the winning region of player $s$ in
$G_c$ (resp.\ $G_{c'}$).

Now let $\sigma$ be
a 
%pure, memoryless 
strategy for player $0$ that is winning on $W_0$ in
$G_c$.
We use that plays that begin in $W_0$ and are consistent with 
$\sigma$ and any 
%pure, memoryless 
strategy $\pi$ of player $1$ are
decided by their periodic suffix~--~which forms a simple cycle as both
strategies are memoryless. As
$c\sim c'$, that decision is the same in both parity games. So $W_o$
is contained in $W_0'$ and $\sigma$ is winning on $W_0$ in game
$G_{c'}$ as well.

A symmetric argument for the winning region $W_1$ and a
$\pi$ for player $1$ that is winning on $W_1$ in $G_{c'}$ then proves the claim
by the determinacy of parity games.

To show that $\equiv$ is contained in $\sim$, let $c\equiv c'$ be
given.
We construct, for each
simple cycle $C$, a 1-player parity game (so one of $V_0$ and $V_1$ is empty)
which is controlled by the player that matches the parity of the $c$-color of $C$. From $c\sim
c'$ is then follows that the $c'$-color of $C$ also has that parity.
(A full proof is contained in the appendix.)\qed

Next, we define the relevant measure of descriptive complexity, which
will also serve as a measure of precision for the static analyses we
will develop.

\begin{definition}
\begin{compactenum}
\item For colored arena $(V,E,c)$, its index $\msr c$ is $\max _{v\in
    V} c(v)$.  
\item The Rabin index $\rabinindex c$ of colored arena $(V,E,c)$ is
  $\min \{ \msr {c'}\mid c\equiv c'\}$.
\item The Rabin index of parity game $(V,V_0,V_1,E,c)$ is $\rabinindex
  c$ for $(V,E,c)$.
\end{compactenum}
\end{definition}

The index $\msr c$ reflects the maximal color occurring in $c$. So for
a coloring function $c\colon V\to \Nat$ on $(V,E)$, its Rabin 
index is the minimal possible maximal color in a coloring function
that is equivalent to $c$. This definition applies to colored arenas
and parity games alike. 

As an aside, is $\msr c$ a good measure, given that $\msr {c+n} = n +
\msr c$ for 
$c+n$ with $(c+n)(v) = c(v) + n$ when $n$ is even? And given that $c$
may have large color gaps?
Fortunately, this is not a concern for the Rabin index of $c$. 
This is so as
for all
$c'\equiv c$ with $\msr {c'} = \rabinindex c$ we know that the minimal
color of $c'$ is at most $1$ and that $c'$ has no color gaps~--~due to
the minimality of the Rabin index.

Intuitively, in order to prove that $\rabinindex c<k$ for some $k>0$
one has to produce a coloring $c'$ and show that all simple
cycles in the graph have the same color under $c$ and $c'$.
As we will see below, 
deciding for a given colored arena $(V,E,c)$ whether $\rabinindex c$
is at least $k$ 
is NP-hard for fixed $k\geq 2$.

Next, we present an algorithm that computes a coloring function
which witnesses the Rabin index of a given $c$. 

\section{Computing the Rabin Index}
\label{section:algorithm}
We now discuss our algorithm $\ccc$, shown in Figure~\ref{fig:ccc}. 
It takes a coloring function as input and outputs an
equivalent one whose index is the Rabin index of the input.
Formally, $\ccc$ computes a coloring function $c'$ with $c\equiv c'$ and
where there is no $c\equiv c''$ with $\msr {c''} <
\msr {c'}$. Then, $\rabinindex c = \msr {c'}$ by definition.
\begin{figure}[bt]
\begin{center}
{\small
\begin{alltt}
rabin(\(V\!,E,c\)) \{
  rank = \(\sum\sb{v\in\!\!\!\! V} \!\!\!\! c(v)\);
  do \{
    cache = rank;
    cycle(); pop();
    rank = \(\sum\sb{v\in\!\!\!\! V} \!\!\!\! c(v)\);
  \} while (cache != rank)
  return \(c\);
\} 

cycle() \{
  sort V in ascending c-color ordering \(v\sb{1}\),\(v\sb{2}\),...,\(v\sb{n}\);
  for (\(i\)=\(1\)..\(n\)) \{
    \(j\) = getAnchor(\(v\sb{i}\));
    if (\(j\) == \(-1\)) \{ \(c(v\sb{i})\) = \(c(v\sb{i})\mathop{\%}2\); \}
    else \{ c(\(v\sb{i}\)) = \(j+1\); \}
  \}
\}

getAnchor(\(v\sb{i}\)) \{
  for (\(\gamma\) = \(c(v\sb{i})-1\) \hbox{ down to }\((c(v\sb{i})-1)\mathop{\%}2\); \hbox{ step size }\(2\)) \{
     if (\(\exists \hbox{simple cycle }C\hbox{ with color }\gamma\hbox{ through }\) \(v\sb{i}\)) \{ return \(\gamma\); \}
  \}
  return \(-1\);
\}

pop() \{
  \(m\) = \(\max\{ c(v)\mid v\in\! V\}\); 
  while (not \(\exists\) simple cycle \(C\) with color \(m\)) \{
    for (\(v\) in \(\{ w\in V\mid c(w) = m\}\)) \{ \(c(v) = m - 1\); \}
    \(m = m - 1\);
  \}
\}
\end{alltt}
}
\end{center}
\vspace*{-5mm}
\caption{Algorithm $\ccc$ which relies on methods \cycle, \getanchor, and
  \pop.\label{fig:ccc}}
\vspace*{-5mm}
\end{figure}

Algorithm $\ccc$ uses a standard iteration pattern based on a rank
function which sums up all colors of all nodes. In each iteration, two
methods are called:

\begin{compactitem}
\item  \cycle\    analyzes the cyclic structure of
$(V,E)$ and so reduces colors of nodes

\item \pop\  repeatedly lowers all occurrences of maximal colors by
  $1$ until there is a simple cycle whose color is a maximal color.
\end{compactitem}

These iterations proceed until neither \cycle\ nor \pop\ has an effect on
the coloring function.  Method \cycle\ 
first sorts all nodes of $(V,E,c)$ in ascending color values for
$c$. It then processes each node $v_i$ in that ascending order. For
each node $v_i$ it calls \getanchor\ to find (if possible) a maximal
``anchor'' for $v_i$.

If \getanchor\ returns $-1$, then $v_i$ has no anchor as all simple cycles through $v_i$ have color $c(v_i)$. Therefore, it is sound to change $c(v_i)$ to its parity. Otherwise, \getanchor\  returns an index $j$ to an ``anchor'' node that is maximal in that

\begin{compactitem}
\item there is a simple cycle $C$ through $v_i$ whose color $j$ is smaller and of different parity than that of $v_i$, and

\item for all simple cycles $C'$ through $v_i$,
 either they have a color that has the same parity as the color of $v_i$ or they have a color that is less than or equal to $j$.
\end{compactitem}
\noindent
A node on this simple cycle $C$ with color $j$ is thus a maximal
anchor for node $v_i$. Method \cycle\ therefore resets $c(v_i)$
to $j + 1$.

The idea behind \pop\  is that one can safely lower maximal
color $m$ to $m-1$ if there is no simple cycle whose color is $m$. For
then all occurrences of $m$ are dominated by smaller colors on simple cycles.

We now prove the soundness of our algorithm $\ccc$.

\begin{lemma}
\label{lemma:cccsound}
Let $(V,E,c)$ be a given colored arena and let $c'$ be the coloring function that is
returned by the call $\ccc(V,E,c)$. Then $c\equiv c'$ holds.
\end{lemma}

We show some example runs of $\ccc$, starting with a detailed worked example, for the parity game in
Figure~\ref{fig:pg}. Let the initial sort of $\cycle$ be $v_3v_4v_2v_0v_1$. Then $\cycle$ changes no colors at $v_3$ (as the anchor of $v_3$ is $-1$), at $v_4$ (as the anchor of $v_4$ is $1$ due to simple cycle $v_4v_3$), at $v_2$ (as the anchor of $v_2$ is $1$ due to simple cycle $v_2v_1v_0v_4v_3$), but changes $c(v_0)$ to $1$ (as the anchor of $v_0$ is $-1$). Also, $c(v_1)$ won't change (as the anchor of $v_1$ is $2$ due to simple cycle $v_1v_2$). 

Then $\pop$ changes $c(v_1)$ to $2$ (as there is no simple cycle with
color $3$). Let the sort of the second call to $\cycle$ be
$v_0v_3v_1v_2v_4$. Then the corresponding list of anchor values is
$-1,-1,1,1,1$ and so $\cycle$ changes no colors. Therefore, the second
call to $\pop$ changes no colors either. Thus the overall effect of
$\ccc$ was to lower the index from $3$ to $2$ by lowering $c(v_1)$ to
$2$.

\begin{figure}[t]
\begin{center}
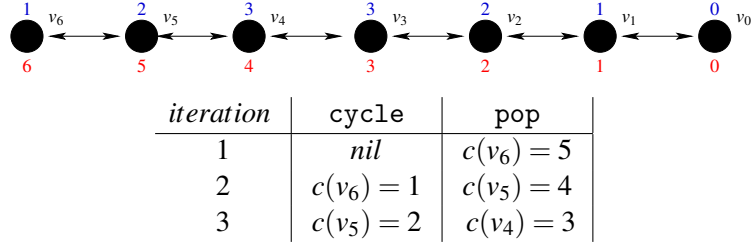
\end{center}
\[
\begin{array}{c|c|c|}
iteration\ & \ \cycle\  & \ \pop\  \\
\hline
1 & \ nil\  & \ c(v_6) = 5\  \\
2 & \ c(v_6) = 1\  & \ c(v_5) = 4\  \\
3 & \ c(v_5) = 2\  & \ c(v_4) = 3
\end{array}
\]
\vspace*{-5mm}
\caption{Colored arena $(V,E,c)$ and table showing effects of
  iterations in $\ccc(V,E,c)$\label{fig:severaliterations}}
%\vspace*{-5mm}
\end{figure}

As a second example, in Figure~\ref{fig:severaliterations}, we see a colored
arena with $c(v_i) = i$ (in red/bottom), the output $\ccc (V,E,c)$ (in
blue/top),
and a table showing how the coloring function changes through
repeated calls to \cycle\  and \pop. Each iteration of $\ccc$ reduces
the measure $\mu(c)$ by $1$. This illustrates that the number of iterations of $\ccc$ is unbounded 
in general.

\begin{figure}[t]
\begin{center}
\subfigure[Coloring functions $c$ and $c'$ give rise to the same winning regions, but not the same winning strategies. Thus $c\not\equiv c'$\label{fig:strategies}]{
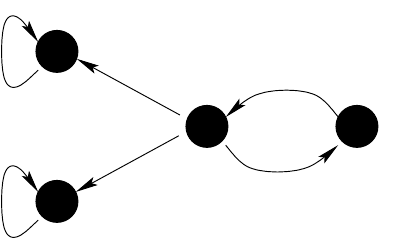
}
%\qquad
\subfigure[Coloring function $c$ has Rabin index $2$, witnessed by $c'$ \label{fig:linear}]{
\hspace*{0.5in}
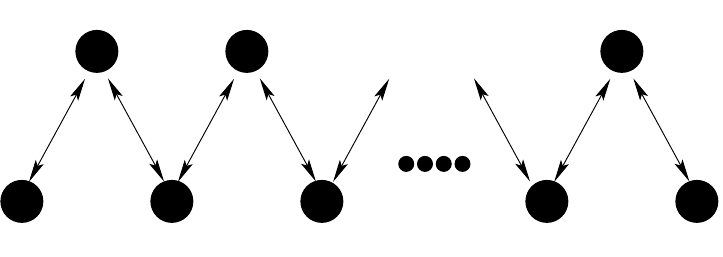
}
\end{center}
\vspace*{-5mm}
\caption{Two coloring functions $c$ (in red/bottom) and $c'$ (in blue/top) on the same game} 
\end{figure}

We note that $\equiv$ cannot be captured by just insisting that the
winning regions of all abstracted parity games be the same. In
Figure~\ref{fig:strategies}, we see a colored arena with two coloring
functions $c$ (in red/bottom) and $c'$ (in blue/top). 
The player who owns node
$v$ will win all nodes as she chooses between $z$ or $o$ the node that
has her parity.
So $c$ and $c'$ are equivalent in that
they always give rise to the same winning regions. 
But if $v$ is owned
by player $1$, she has a winning strategy for $c'$ (move from $v$
to $w$) that is not winning for $c$.

In Figure~\ref{fig:linear}, colored arena $(V,E,c)$ has odd index $n$
and Rabin index $2$. Although there are cycles from all nodes with
color $n$, e.g., to the node with color $n-1$, there are no
\emph{simple} such cycles. So all colors reduce to their
parity.

Now we can prove that algorithm $\ccc$ is basically as precise as it
could be. First, we state and prove an auxilliary lemma which provides sufficient conditions for a coloring function $c$ to have its index $\msr c$ as its Rabin index $\rabinindex c$.
Then we show that the output of $\ccc$ meets these conditions.

\begin{lemma}
\label{lemma:c}
Let $(V,E,c)$ be a colored arena where

\begin{compactenum}
\item there is a simple cycle in $(V,E)$ whose color is the maximal
  one of $c$

\item for all $v$ in $V$ with $c(v) > 1$, node $v$ is on a simple cycle $C$
 with color $c(v)-1$.
\end{compactenum}
\noindent
Then there is no $c'$ with $c\equiv c'$ and $\msr {c'} < \msr c$. And
so $\msr c$ equals $\rabinindex c$.
\end{lemma}

\ourproof {}
Let $k$ be the maximal color of $c$ and consider an arbitrary $c'$ with
$c\equiv c'$. 

{\bf Proof by contradiction:} Let the maximal color $k'$ of $c'$
satisfy $k' < k$. 
By the first assumption, there is a  simple cycle $C_0$
whose $c$-color is
$k$. Since $k' < k$ and $c\equiv c'$, we know that the $c'$-color of
$C_0$ can be at most $k-2$. Let $v_0$ be a node on $C_0$ such that
$c'(v_0)$ is the $c'$-color of $C_0$. Then $c'(v_0)\leq k-2$. 
As all nodes on $C_0$ have
$c$-color $k$, we have also $c(v_0)\geq k$. For $k < 2$,
then $c'(v_0)\leq k-2$ gives us a contradiction $c'(v_0) < 0$. It thus remains to
consider the case when $k\geq 2$.

By the second assumption, there is some simple cycle $C_1$ through
$v_0$ such that the color of $C_1$ is $k-1$. In particular, there is
some node $v_0'$ in $C_1$ with color $k-1$. But $k-1$ cannot be the
color of $C_1$ with respect to $c'$ since $v_0$ is on $C_1$ and
$c'(v_0)\leq k-2$. Since $c\equiv c'$, the $c'$-color of $C_1$ is
therefore at most $k-3$. So there is some $v_1$ on $C_1$ 
such that $c'(v_1)\leq k-3 < k - 1 \leq c(v_1)$.

If $c(v_1) > 1$, we repeat the above argument at node $v_1$ to
construct a simple cycle $C_2$ through $v_1$ with color
$c(v_1)-1$. Again, there then have to be nodes $v_1'$ and $v_2$ on
$C_2$ such that the color $c'(v_1')$ is the $c'$-color of $C_2$, and
such that $c'(v_2)\leq k - 4 < k - 2 \leq c(v_2)$ holds.

We can repeat the above argument to construct simple cycles
$C_0,C_1,C_2,\dots$ and nodes $v_0,v_0',v_1,v_1',v_2,v_2',\dots$ such
that $c'(v_j) \leq k - j -2 < k - j \leq c(v_j)$ until $k - j\leq
c(v_j) \leq 1$. But then 
$c'(v_j) \leq k - j - 2 \leq 1 - 2 = -1$,  a contradiction.
\qed

We now show that the output of $\ccc$ satisfies the assumptions of
Lemma~\ref{lemma:c}. Since \ccc\ is sound for $\equiv$,
we therefore infer that it computes a coloring function whose maximal color
equals the Rabin index of its input coloring function.

\begin{theorem}
Let $(V,E,c)$ be a colored arena. And let $c^*$ be the output of the call $\ccc(V,E,c)$. Then $c\equiv c^*$ and
$\msr {c^*}$ is the Rabin index of $c$.
\end{theorem}

\ourproof {}
By Lemma~\ref{lemma:cccsound}, we have $c\equiv c^*$. Since $\equiv$
is clearly transitive, it suffices to show that there is no $c'$ with
$c^*\equiv c'$ and $\msr {c'} < \msr {c^*}$. By Lemma~\ref{lemma:c},
it therefore suffices to establish the two assumptions of that lemma
for $c^*$. 
As $c^*$ is returned by $\ccc$ neither \cycle\ nor \pop\ have an
effect on it.

The first assumption of Lemma~\ref{lemma:c} is therefore true since
\pop\  has no effect 
on $c^*$ and so there must be a simple cycle in $(V,E)$ whose color is
the maximal one in $c$. 
This also applies to the case when
$c^*$ has only one color, as $(V,E)$ has to contain cycles since it is
finite and all nodes have outgoing edges.

As for the second assumption, let by way of contradiction there be
some node $v$ with $c^*(v) > 1$ and no simple
cycle through $v$ with color $c^*(v)-1$. Then \cycle\  would have
an effect on $c^*(v)$ and would lower it, a contradiction.
\qed

\section{Complexity}
\label{section:complexity}

We now discuss the complexity of algorithm $\ccc$ and of the decision problems associated with the Rabin index. 
We turn to the complexity of $\ccc$ first. 

Let us assume that we have an oracle that checks for
the existence of simple cycles. Then the computation of  $\ccc$ is
efficient modulo polynomially many calls (in the size of the game) to that oracle. Since
deciding whether a simple cycle exists between
two nodes in a directed graph is NP-complete (see e.g.\
\cite{EvenIS76,Fortune80}), we infer that  $\ccc$ can be implemented to run in exponential time.
 
Next, we study the complexity of deciding the value of the Rabin index.
We can exploit the NP-hardness of simple cycle detection to show that
the natural decision problem for the Rabin index, whether $\rabinindex
c$ is at least $k$, is NP-hard for fixed
$k\geq 2$. In contrast, for $k=1$, we show that this problem is in P.

\begin{theorem}
Deciding whether the Rabin index of a colored arena $(V,E,c)$ is
at least $k$ is NP-hard 
for every fixed $k\geq 2$, and is in P for $k = 1$.
\end{theorem}

\ourproof {}
First consider the case when $k\geq 2$.
We use the fact that deciding whether there is a simple cycle through
nodes $s\not=t$ in a directed graph $(V,E)$ is NP-complete (see e.g.\
\cite{Fortune80}). 
Without loss of generality, for all $v$ in $V$ there is some $w$ in $V$ with
$(v,w)$ in $E$ (we can add $(v,v)$ to $E$ otherwise). Our
hardness reduction uses a colored arena $(V',E',c)$,  depicted in Figure~\ref{fig:hardness}, which we now describe:
\begin{figure}[tb]
\begin{center}
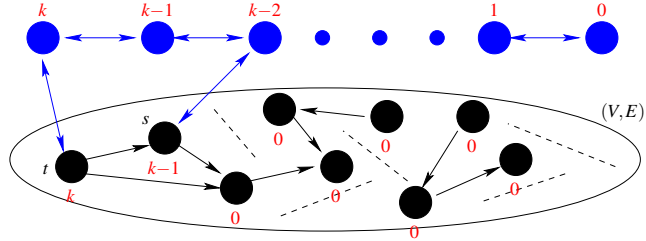
\end{center}
\vspace*{-5mm}
\caption{Construction for NP-hardness of deciding whether $\rabinindex
  c \geq k$ for $k\geq 2$\label{fig:hardness}}
\vspace*{-5mm}
\end{figure}

We color $s$ with $k-1$ and $t$ with $k$, and color all remaining
nodes of $V$ with $0$. Then we add $k+1$ many new nodes (shown in blue/top in
the figure) to that graph that form a ``spine'' of descending
colors from $k$ down to $0$, connected by simple cycles. Crucially, we
also add a simple cycle 
between $t$ and that new $k$ node, and between $s$ and the new $k-2$
node. 

We claim that the Rabin index of $(V',E',c)$ is at least $k$
iff there is a 
simple cycle through $s$ and $t$ in the original directed graph $(V,E)$.

{\bf 1.} Let there be a simple cycle through $s$ and $t$ in 
$(V,E)$. Since there is a simple cycle between
$s$ and the new $k-2$ node, \cycle\
does not change the color at $s$. As there is a simple cycle through $s$ and
$t$, method \cycle\ also does not change the color at
$t$.
Clearly, no colors on the spine can be changed by \cycle. Since there
is a simple cycle between $t$ and the new $k$ node, method \pop\ also
does not change colors. But then the Rabin index of $c$ is $k$ and so
at least $k$.

{\bf 2.} Conversely, 
assume that there is no simple cycle through $s$ and $t$ in the
original graph $(V,E)$.
It follows that the anchor $j$ of $t$ has value $0$ or, if $k$ is even,
has value $-1$.
In this case, \cycle\ changes the color at $t$ to the parity of $k$.
Then, \pop\ reduces the color of the remaining node colored $k$ to $k-1$.
Thus, it cannot be the case that the Rabin index of $c$ is at least $k$.

This therefore proves the claim.
Second, consider the case when $k=1$.
Deciding whether $\rabinindex c$ is at least $1$ amounts to checking whether
$c\equiv \vec 0$ where $\vec 0(v) = 0$ for all $v$ in $V$. 
This is the case iff all simple cycles in $(V,E,c)$ have even
$c$-parity. 
But that is the case iff all cycles in $(V,E,c)$ have even $c$-parity.

To see this, note that the ``if'' part is true as simple cycles
are cycles. 
As for the ``only if'' part, this is true since if there
were a cycle $C$ with odd $c$-parity, then some node  
$v$ on that
cycle would have to have that minimal $c$-color, but $v$ would then be
on some simple cycle whose edges all belong to $C$.

Finally, checking whether all cycles in $(V,E,c)$ have even $c$-parity is in P.
\qed

The decision
problem of whether $\rabinindex c = 1$ cannot be in NP, unless NP equals coNP. Otherwise,
the decision problem of
whether $\rabinindex c \leq 1$ would also be in NP, since we can
decide in P whether $\rabinindex c = 0$ and since NP is closed under
unions. 
But then the complement decision
problem of whether $\rabinindex c \geq 2$ would be in coNP, and we
have shown it to be NP-hard already. Therefore, all problems in NP would reduce to
this problem and so be in coNP as well, a contradiction.

We now discuss an efficient version of $\ccc$ which
replaces oracle calls for simple cycle detection with
calls for over-approximating cycle detection. 

\section{Abstract Rabin index}
\label{section:static}
We now discuss an efficient version of $\ccc$ which
replaces oracle calls for simple cycle detection with
over-approximating cycle detection. 
In fact, this static analysis computes an abstract Rabin index, whose
definition is based on an abstract version of the equivalence relation
$\equiv$. We define these notions formally. 

\begin{definition}
\begin{compactenum}
\item Let $\cccai$ be  $\ccc$ where all existential
quantifications over simple cycles are replaced with existential
quantifications over cycles.

\item Let $(V,E)$ be a directed graph and $c,c'\colon V\to \Nat$ two
  coloring functions. Then:

  \begin{compactenum}
  \item $c\equivai c'$ iff
for all cycles $C$, the parities of their $c$- and $c'$-colors are
equal. 

   \item The abstract Rabin index
$\rabinindexai c$ of $(V,E,c)$ is $\min \{ \msr {c'} \mid
c\equivai c'\}$.
  \end{compactenum}
\end{compactenum}
\end{definition}

\begin{figure}[bt]
\begin{center}
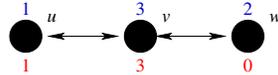
\end{center}
\vspace*{-5mm}
\caption{Coloring functions $c$ (blue/top) and $c'$ (red/bottom) with
  $c\equiv c'$ but $c\not\equivai c'$\label{fig:aidiff}}
\vspace*{-5mm}
\end{figure}

Thus $\cccai$ uses the set of cycles in $(V,E)$ to overapproximate the
set of simple cycles  in $(V,E)$. In particular, $c\equivai c'$ implies $c\equiv c'$ but not the other
way around, as can be seen in the example in Figure~\ref{fig:aidiff}.

In that example, we have $c\equiv c'$ since all simple cycles have the same parity of color with respect to $c$ and $c'$. But there is a cycle
that reaches all three nodes and which has odd color for $c$ and even color for $c'$. Thus, $c\not\equivai c'$ follows.

We now show that the overapproximation $\cccai$ of $\ccc$ is sound in
that its output coloring function is equivalent to its input
coloring function. Below, in Theorem~\ref{theorem:riai}, we further
show that this output yields an abstract Rabin index.

\begin{lemma}
\label{lemma:soundai}
Let $(V,E,c)$ be a colored arena and let $\cccai (V,E,c)$ return $c'$. Then $c\equivai c'$ and $\msr
{c'}\geq \rabinindex c$. 
\end{lemma}

To prove this lemma, it suffices to show $c\equivai c'$, as  $c\equiv c'$ follows from that, and then this in turn implies $\msr {c'}\geq \rabinindex c$ by the
definition of the 
Rabin index.

Note that the definition of $\equivai$ is like the characterization of
$\equiv$ in Proposition~\ref{prop:equiv},
except that the universal quantification over simple cycles 
is
being replaced by a universal quantification over cycles for
$\equivai$.
In proving Lemma~\ref{lemma:soundai}, we can thus reuse the proof for Lemma~\ref{lemma:cccsound} where
we replace $\equiv$ with $\equivai$, $\ccc$ with $\cccai$,  and ``simple cycle'' with ``cycle''
throughout in that proof. 

We can now adapt the results for $\ccc$ to this
abstract setting.

\begin{lemma}
\label{lemma:a}
Let $(V,E,c)$ be a colored arena where

\begin{compactenum}
\item there is a cycle in $(V,E)$ whose color is the maximal
  one of $c$

\item for all $v$ in $V$ with $c(v) > 1$, node $v$ is on a cycle $C$
 with color $c(v)-1$.
\end{compactenum}
\noindent
Then there is no $c'$ with $c\equivai c'$ and $\msr {c'} < \msr c$, and so $\msr c = \rabinindexai c$.
\end{lemma}

Similary to the case for algorithm $\ccc$, we now show that the output of $\cccai$ satisfies the assumptions of
Lemma~\ref{lemma:a}. Since algorithm $\cccai$ is sound for $\equivai$,
we therefore infer that it computes coloring functions whose maximal color
equals the abstract Rabin index of their input coloring function.

\begin{theorem}
\label{theorem:riai}
Let $(V,E,c)$ be a colored arena. And let $c^*$ be the output of the call $\cccai(V,E,c)$. Then $c\equivai c^*$ and
$\msr {c^*}$ is the abstract Rabin index $\rabinindexai c$.
\end{theorem}

We now study the sets of parity games  whose abstract
Rabin index is below a fixed bound. We define these sets formally.

\begin{definition}
Let $\mathcal P_k^\alpha$ be the set of parity games
  $(V,V_0,V_1,E,c)$ with $\rabinindexai c < k$.
\end{definition}

We can now show that parity games in these sets are efficiently solvable, also in
the sense that membership in such a set is efficiently decidable.

\begin{theorem}
Let $k\geq 1$ be fixed. All parity games in $\mathcal P_k^\alpha$
can be solved in polynomial time. Moreover, membership in $\mathcal
P_k^\alpha$ can be decided in polynomial time.
\end{theorem}

\ourproof {}
For each parity game $(V,V_0,V_1,E,c)$ in $\mathcal P_k^\alpha$, we first run $\cccai$ on it, which runs in polynomial time. By definition of $\mathcal P_k^\alpha$, the output coloring function $c^*$ has index $< k$.
Then we solve the parity game $(V,V_0,V_1,E,c^*)$, which we can do in
polynomial time as the index is bounded by $k$. But that solution is
also one for $(V,V_0,V_1,E,c)$ since $c\equivai c^*$ by Lemma~\ref{lemma:soundai}, and so $c\equiv c^*$ as well.

That the membership test is polynomial in the running time can be seen
as follows: for coloring function $c$, compute $c' = \cccai (V,E,c)$ and return {\tt true}
if $\msr {c'} < k$ and return {\tt false} otherwise; this is correct by Theorem~\ref{theorem:riai}.
\qed

We note that algorithm $\cccai$ is precise for colored arenas
$A=(V,E,c)$ with Rabin index $0$. These are colored arenas that have
only simple cycles with even color. 
Since a colored arena has a cycle with odd color iff it has a simple
cycle with odd color, $\cccai$ correctly reduces all colors to $0$
for such arenas. 

For Rabin index $1$, the situation is more subtle. We cannot expect
$\cccai$ to always be precise, as the decision problem for
$\rabinindex c \geq 2$ is NP-hard. Algorithm $\cccai$ will correctly
compute Rabin index $1$ for all those arenas that do not have a simple
cycle with even color. But for $c$ from Figure~\ref{fig:aidiff}, e.g.,
algorithm $\cccai$ does not change $c$ with index $3$, although the Rabin index
of $c$ is $1$. 

\section{Experimental results}
\label{section:experiments}
We now provide some experimental results.
Our objective is
to compare the effectiveness of color
compression of $\cccai$ to a known color compression algorithm (called
static compression), to observe the performance improvement 
in solving compressed games using Zielonka's parity game solver
\cite{Zie98}, and to get a feel for how much the abstract
Rabin index reduces the index of random and non-random games.  

Our implementation is written in Scala and realizes all game elements
as objects to simplify implementation.
Our main interest is in descriptive complexity measures and relative
computation time.

We programed algorithm $\ccc$ with simple cycle detection
reduced to incremental SAT solving. This did not
scale to graphs with more than $40$ nodes.
But for those games for which we could compute
the Rabin index, $\cccai(V,E,c)$ often computed the Rabin index
$\rabinindex c$ or did get very close to it.

Our implementation of $\cccai$ reduced cycle detection to the
decomposition of the graph into strongly connected components, using
Tarjan's algorithm (which is linear in the number of edges). The rank function is only needed for complexity and termination analysis, we replaced it with Booleans that flag
whether \cycle\ or \pop\ had an effect.

\begin{figure}[tb]
\begin{center}
{\small
\begin{tabular}{|l||r|r|r||r|r||r||r|r|r|}
\hline
Game Type & $\mu(c)$ & $\mu(s(c))$ & $\rabinindexai c$ & S & R & \#I & Sol & Sol.S & Sol.R \\ \hline\hline
{\tt Clique}[100] & 100 & 100 & 99  & 0.08 & 388.93& 2 & 13.23 & 13.06 & 13.01 \\ \hline 
{\tt Ladder}[100] & 2 & 2 & 2 & 0.11 & 8.93& 1  & 1.87 & 1.66 & 1.68 \\ \hline 
{\tt Jurdzi\'{n}ski}[5 10] & 12 & 12 & 11 & 0.09 & 44.25 & 2 & 76.98 & 76.94 & 76.38 \\ \hline 
{\tt Recursive Ladder}[15] & 48 & 46 & 16 & 0.04 & 10.46 & 2 & 310.21 & 309.21 & 174.91 \\ \hline 
{\tt Strategy Impr}[8] & 237 & 181 & 9 & 0.10 & 54.01 & 2 & 194.96 & 45.46 & 8.99 \\ \hline 
{\tt Model Checker Ladder}[100] & 200 & 200 & 0 & 0.14 & 141.95 & 2 & 30.90 & 30.49 & 0.62 \\ \hline 
{\tt Tower of Hanoi}[5] & 2 & 2 & 1 & 0.46 & 261.10& 2  & 29.43 & 29.61 & 45.41 \\ \hline 
\end{tabular}
}
\end{center}
\vspace*{-5mm}
\caption{Indices and average times (in $ms$) for $100$ runs for game
  types named in first column.
Next three columns: original, statically compressed, and
$\cccai$-compressed index.
Next three columns: times of static and $\cccai$-compression, and the number
of iterations within $\cccai$. Last three
columns: Times of solving the original, statically compressed, and $\cccai$-compressed games with Zielonka's solver}
\label{fig:results} 
\vspace*{-5mm}
\end{figure}

The standard static compression algorithm 
simply removes gaps between colors, e.g.\ a set of colors
$\{0,3,4,5,6,8\}$ is being compressed to $\{0,1,2,3,4\}$. 
Below, we write $s(c)$ for the statically compressed version of
coloring function $c$.

The experiments are conducted on non-random and random games
separately. Each run of the experiments generates a parity game $G$ =
($V, V_0, V_1, E, c$) of a selected configuration. Static compression
and $\cccai$ are performed on these games. 
We report the time taken to execute static compression and $\cccai$,
as well as the number of iterations that $\cccai$ runs until \cycle\
and \pop\ have no effect, i.e.\ the number of iterations needed for
$\mu(c)$ to reach $\rabinindexai c$.  Finally, we record the
wall-clock time required to solve original, 
statically compressed, and $\cccai$-compressed games, using Zielonka's solver \cite{Zie98}.

We use PGSolver to generate non-random games, detailed descriptions on
these games can be found in \cite{Friedmann10}. Each row in
Figure~\ref{fig:results} shows the average statistics from 100 runs of
the experiments on corresponding non-random game. We see that $\cccai$
has significantly reduced the indices of 
{\tt  Recursive Ladder}, {\tt Strategy Impr}, and {\tt Model Checker
  Ladder}, where $\rabinindexai c$ is 0\% to 
35\% of the index
$\mu(s(c))$ of the statically compressed coloring function. 

Applying $\cccai$ improves performance of solvers.
For all 
three game types, we observe 
44\% to 98\%
in solver time reduction between solving statically compressed and
$\cccai$-compressed games.

The time required to perform static compression is low compared to the
time needed for $\cccai$-compression, but $\cccai$-compression followed by solving the game is still faster than solving the original game for 
{\tt Recursive Ladder}.

Games {\tt Ladder} and {\tt Tower of Hanoi} have very low indices and
their colors cannot be compressed further.
Method \cycle\ has no effect on {\tt Clique} games, but \pop\ manages to reduce its index by $1$.
\begin{figure}[tb]
\begin{center}
\begin{tabular}{|l||r|r|r||r|r||r||r|r|r|}
\hline
Game Configs & $\mu(c)$ & $\mu(s(c))$ & $\rabinindexai c$ & S &  R & \#I  & Sol & Sol.S & Sol.R \\ \hline\hline
100/1/20/100 & 99.16 & 45.34 & 35.97& 0.48 & 57.04 & 2.05  & 6.71 & 5.21 & 4.84 \\ \hline 
200/1/40/200 & 198.97 & 91.91 & 80.29& 0.12 & 441.29 & 2.03  & 12.40 & 11.49 & 11.43 \\ \hline 
400/1/80/400 & 399.28 & 184.34 & 172.30 & 0.24 & 4337.04 & 2.10 & 42.78 & 40.62 & 40.58 \\ \hline 
800/1/160/800 & 799.08 & 369.76 & 355.67& 0.47 & 47241.70 & 2.05  & 181.73 & 173.59 & 173.83 \\ \hline 
1000/1/200/1000 & 999.14 & 462.48 & 447.37& 0.59 & 106332.96 & 2.05  & 296.53 & 281.60 & 281.70 \\ \hline 
\end{tabular}
\end{center}
\vspace*{-5mm}
\caption{
\label{fig:random-results} 
Indices and average times (in $ms$) for $100$ runs of random
  games of various configurations listed in the first column. Next
  three columns: 
  average original, statically compressed, and
$\cccai$-compressed indices. The remaining columns are as in
Figure~\ref{fig:results}} 
\vspace*{-5mm}
\end{figure}

We now discuss our experimental results on random games. The notation used to describe randomly generated parity games is $xx/yy/zz/cc$, where $xx$ is the number of nodes (node ownership is determined by a fair coin flip for each node independently), with between $yy$ to $zz$ out-going edges for each node, and with colors at nodes chosen at random from $\{0,\dots,cc\}$. Also, the games used in the experiments have $1$ as the minimum number of out-going edges. This means that the nodes have no dead-ends. We also disallow self-loops (no $(v,v)$ in $E$). 

Figure~\ref{fig:random-results} shows the average statistics of $100$
runs of experiments on five selected game configurations. (Our
experiments on larger games are consistent with the data reported
here, and so not reported here.) The results indicate that
static compression is effective in reducing the colors for 
randomly generated games, it achieves 
around 54\% index reduction for all game types. The $\cccai$-compression achieves further 
3\% to 21\%
 reduction. Due to the relatively small index reduction by $\cccai$, we do not see much improvement in solving $\cccai$-compressed games over solving statically-compressed ones. In addition, $\cccai$ reduces $\mu(c)$ to $\rabinindexai c$ in one iteration for all of the randomly generated games $G$. 

The results in Figure~\ref{fig:random-results} show that these games take an average of more than 2 $\cccai$ iterations. 
This indicates that certain game structure, such as the one found in the game in Figure~\ref{fig:severaliterations}, is present in our randomly generated games

The experimental results show that $\cccai$ is able to reduce the
indices of parity games significantly and quickly, for certain
structure such as {\tt Recursive Ladder}. 
Hence it effectively improves the overall solver performance for those games. 

However, algorithm $\cccai$ has a negative effect on the overall performance for other non-random games and experimented random games, when we consider $\cccai$-compression time plus solver time.

\hide{study alternative forms of Rabin index, e.g.\ for the measure
  that uses a lexicographical ordering over $\Nat\times \Nat$ where
  the first number represents $\mu (c)$ and the second one the sum of
  all colors at all nodes}

\section{Related work}
\label{section:relatedwork}

\begin{figure}[hbt]
\begin{center}
{\small
\begin{alltt}
Rabin\(\sb{a}\)(\(V,E,c\)) \{
  define a new colouring function \(c'\) for \((V,E,c)\);
  reduce(\(V,E,c,c'\));
  return \(c'\);
\}
reduce(\(V,E,c,c'\)) \{
  i = 0; decompose \((V,E)\) into maximal \(SCCs\);
  for (\(R \in SCCs\))\{
    if (\(\pi(R)\) == 0) m = 0;
    else \{
      \(R'\) = \{\(v \in R \mid c(v) \neq \pi(R)\)\}; m = reduce(\(R',E|\sb{R'},c|\sb{R'},c'|\sb{R'}\));
      if (\(\pi(R)\) - m is odd) m = m + 1;
    \}
    for (\(v \in \{v\in R \mid c(v) = \pi(R)\}\)) 
      \(c'(v)\) = m;
    i = max\{i, m\};
  \}
  return i;
\}
\end{alltt}
}
\end{center}
\vspace*{-5mm}
\caption{Algorithm to compute Rabin index \cite{CartonM99} for a
  parity automaton $A$ = ($V$, $E$, $c$), where $R\subseteq V$,
  $\pi(R)$ = max\{$c(v) \mid v \in R$\}, 
  $E|_{R}$ is $E$ with restriction to nodes in $R$, and similarly for
  $c|_{R}$.
\label{fig:rabinautomata}}
\vspace*{-3mm}
\end{figure}

Carton and Maceiras develop an algorithm (denoted here $\ccca$) that
computes and minimizes the Rabin index of deterministic parity word
automata \cite{CartonM99}. 
Deterministic parity word automata can be thought of as 1-player
parity games, where the player chooses input letters.
An infinite word can be compared to a strategy with memory for the
player.
The word is accepted if the strategy is winning, that is, if the minimal color to be visited infinitely often is even.
Minimization of the Rabin index should preserve the language of the
automaton or, put in our terms, every winning strategy should remain
to be winning.

The pseudocode of $\ccca$ is shown in Figure~\ref{fig:rabinautomata}. 
Algorithm $\ccca$ constructs the ``coloring dependencies'' of all
states in an automaton arena by decomposing the automaton into maximal
strongly connected components ($SCCs$). 
For each $R$ being a maximal $SCC$, it 
removes the states with the maximal color (and pushes them onto a
stack), then recursively $SCC$ decomposes the remaining arena of
$R$. Eventually, the input arena is reduced to a set of states that
exist in their own respective $SCCs$ (hence do not exist in the same
cycle as each other). 
These states are assigned the minimal colors $m$ (which is $0$ or $1$
depending on their original parities). The algorithm then propagates
the new colour $m$ to the states in the ``layer'' above. Those states
receive a new colour $m$ or $m$ + 1, depending on whether their
original parities equal the parities of the states in the ``layer''
below. In essence, $SCC$ decomposition is used to detect the cycle
dependency of states and this techniques is also used in our
implementation of $\cccai$. 

Our notion of Rabin index is a natural generalization to 2-player
games.
We require that for every pair of strategies $(\sigma,\pi)$, their
outcome should not change.
As mentioned, the weaker notion requiring to preserve winning
strategies of each player separately is not interesting.
Such a Rabin index associates rank 0 with the winning region of
player~0 and 1 with the winning region of player~1.
It can be computed by solving the game.

The transition from 1-player setting to 2-player setting requires a
more elaborate algorithm for computing the Rabin index.
Although presented differently, algorithm $\ccca$ has the same effect
of $\cycle$ in $\cccai$, which approximates the Rabin index.
In our context of 2-player games one has to replace SCC decomposition
(or cycle detection) by simple-cycle detection.
Furthermore, in order to compute the Rabin index of a 2-player game we
have to add the procedure $\pop$.
These two additional components are crucial for the computation of the
Rabin index of games (as shown in this paper).

The differences become crucially important in terms of the
computational complexity and degree of possible color compression in
the setting of parity games.  
Using the colored arena in Figure~\ref{fig:severaliterations} as an
example, $\ccca$ will make no change to the red coloring function,
whereas $\cccai$ reduces its index to $5$ (using $\pop$), and $\ccc$
reduces it even to $3$.

\section{Conclusions}
\label{section:conclusions}

We have provided a descriptive measure of complexity for parity games
that (essentially) measures the number of colors needed in a parity
game if we forget the ownership structure of the game but if we do not
compromise the winning regions or winning strategies by changing its
colors. 

We called this measure the Rabin index of a parity game. We then
studied this concept in depth. By analyzing the structure of simple
cycles in parity games, we arrived at an algorithm that computes this
Rabin index in exponential time. 

Then we studied the complexity of the decision problem of whether the 
Rabin index of a parity game is at least $k$ for some fixed $k >
0$. For $k$ equal to $1$, we saw that this problem is in P, but we
showed NP-hardness of this decision problem for all other values of
$k$. These lower bounds therefore also apply to games that capture
these decision problems in game-theoretic terms.

Next, we asked what happens if our algorithm $\ccc$ abstractly
interprets all detection checks for simple cycles through detection
checks for cycles. The resulting algorithm $\cccai$ was then shown to
run in polynomial time, and to compute an abstract and
sound approximation of the Rabin index. 

Our experiments
were performed on random
and non-random games. We observed that $\cccai$-compression plus
Zielonka's solver \cite{Zie98} 
in some cases speed up solving time.
The combination achieved 29\% and 85\% time reduction
for {\tt Jurdzi\'{n}ski} and {\tt Recursive Ladder} games,
respectively, over solving the original games. But for other game
types and random games, no such reduction was observed.
We also saw that for some structured game types, the abstract Rabin
index is dramatically smaller than the index of the game.

In future work we mean to investigate properties of the measure
$\rabinindexai c - \rabinindex c$. Intuitively, it measures the
difference of the Rabin index based on the structure of cycles with
that based on the structure of simple cycles. 
From Figure~\ref{fig:linear} we already
know that this measure can be arbitrarily large.  

It will also be of interest to study variants of $\rabinindex c$ that
are targeted for specific solvers. For example, the SPM solver in
\cite{Jur00} favors fewer occurrences of odd colors but also favors
lower index. This suggests a measure with a lexicographical order of
the Rabin index followed by an occurrence count of odd colors. 

\bibliographystyle{eptcs}
\bibliography{references}

\end{document}